# Family Structure, Gender, and Subjective Well-being: Effect of Children before and after COVID-19 in Japan

Eiji Yamamura,[a,][*], and Fumio Ohtake[b]

[a]Department of Economics, Seinan Gakuin University, Japan

[b]Center for Infectious Disease Education and Research, Osaka University, Japan, ohtake@cider.osaka-u.ac.jp

*Corresponding author

Department of Economics, Seinan Gakuin University

6-2-92 Sawaraku Nishijin

Fukuoka 814-8511

Japan

Tel.: 81-92-823-4543

Fax.: 81-92-823-2506

Email: yamaei@seinan-gu.ac.jp

Acknowledgments: We would like to thank Editage [http://www.editage.com] for editing and reviewing this manuscript for English language. This study was supported by Fostering Joint International Research B (Grant No.18KK0048) and Grant-in-Aid for Scientific Research S (Grant No. 20H05632) from the Japan Society for the Promotion of Science.

# Abstract

Grandparents were anticipated to participated in grand-rearing. The COVID-19 pandemic had detached grandparents from rearing grandchildren. The research questions of this study were as follows: How does the change in family relations impact the well-being (SWB) of grandparents and parents? Using independently collected individual-level panel data over 2016–2023, we examined how family structure influenced subjective SWB before and after COVID-19. We focused on the effects of children, grandchildren, and their gender on grandparents and parents. We found that compared with the happiness level before COVID-19, (1) granddaughters increased their grandmothers' SWB after COVID-19, (2) both daughters and sons reduced their fathers' SWB after COVID-19, whereas neither daughters nor sons changed their mothers' SWB, and (3) the negative effect of sons reduced substantially if their fathers had younger brothers. Learning from interactions with younger brothers in childhood, fathers could avoid the deterioration of relationships with their sons, even when unexpected events possibly changed the lifestyle of the family and their relationship.

1. Introduction

In 2023, we observed a remarkable decrease in the number of people who wore masks when walking on the street. The phrase "stay home" became a thing of the past. About five years have passed since the emergence of COVID-19 in 2019, and we have entered the post-COVID period. We have seemingly returned to our daily lives before the pandemic. However, the number of births in Japan declined clearly and constantly even before COVID-19 (Figure 1). As suggested in Figure 1, in 2022, the number of births was the first to fall below 800,000 since 1899 when demographics were first collected. An accelerated low birth rate is a critical policy issue especially after COVID-19.

The incentive to have a child can be analyzed by considering the SWB of those who rear the child. The youth are incentivized to have a child if the birth of a child increases their SWB. Their behaviors partly depend on their parents' and grandparents' support for childrearing (Aparicio-Fenoll & Vidal-Fernandez, 2015; Brunello & Yamamura, 2023). However, newborn babies cause a reasonably increased burden of childcare not only for parents (Nomaguchi & Milie, 2003; Stanca, 2012) but also for grandparents (Ahn & Choi, 2019; Brunello & Yamamura, 2023; Del Boca et al., 2018; Ku et al., 2012). Accordingly, having a child reduces parents' (Clark et al., 2008; Nomaguchi & Milie, 2003; Stanca, 2012) and grandparents' SWB (Brunello & Rocco, 2019; Yamamura & Brunello, 2023). The purpose of this study is to examine how childbirth influenced the SWB of grandparents and parents during 2016–2023.

In the short-term analysis, immediately after COVID-19 emerged, the gap in the burden of childrearing increased between mothers and fathers (Yamamura & Tsutsui, 2021a). This might be one of the factors that decreases mothers' SWB (Yamamura & Tsutsui, 2021b). The impact of various life events such as marriage and divorce on SWB is temporary (Clark et al., 2008). If this is true, the perceived cost of childrearing and in turn, the influence of children on SWB returned to the pre-pandemic level as time passed. However, we should consider changes in circumstances owing to the pandemic. Child abuse and domestic violence increased during the pandemic (Karbasi et al., 2022; Kovler et al., 2021; Nguyen, 2021). Once relationships between family members deteriorate, it is difficult to restore them. Little is known about how COVID-19 has changed family relationships and influenced SWB in the long term[1].

To consider family issues, gender differences should be considered because gender identity influences the division of labor between wives and husbands (Akerlof & Kranton, 2000). Global Gender Gap Report 2023 (World Economic Forum 2023) indicate that Japan ranks 125th among 146 countries in The Global Gender Gap Index 2023 rankings. The social status of Japanese women is markedly lower than that of women in other developed countries. This might be associated with the gender gap in the burden of childrearing, which lowers women's incentives to have a child.

After the pandemic began, the gap in the burden of childrearing has widened between mothers and fathers, causing a mother's SWB to decline (Yamamura & Tsutsui, 2021a, 2021b). However, little is known about how COVID-19 altered family relations to influence SWB in the long term. Furthermore, gender-matching between generations within a household is important. For instance, grandmothers have closer relations with their daughters than with sons and daughters in law, enhancing grandparents' support for childrearing, especially in traditional societies (Brunello & Yamamura, 2023). Apart from current parent–child relations, family structure in childhood has a long-term effect on SWB and the perception of childrearing. The SWB is positively associated with close relationships with sisters, but not with brothers (Cicirelli, 1989). An individual's subjective view and characteristics are shaped by sibling structure, such as birth order and the sex of the siblings (Okudaira et al., 2015; Yamamura, 2015). In the estimation, we consider these gender relationships within the household.

Using independently collected individual-level panel data and a simple Fixed Effects (FE) regression, we provide the following findings. Granddaughters increased their grandmothers' SWB after COVID-19, whereas grandchildren had no effect on the grandfathers' SWB. By contrast, both daughters and sons reduced their fathers' SWB after COVID-19, whereas neither daughters nor sons changed their mothers' SWB. If the father had a younger brother, the negative effect of his son reduced, but the negative effect of his daughter did not change.

---

[1] Many studies examined impact of COVID-19 on SWB, for instance, in Japan (Sugawara et al., 2022; Yamamura and Tsutsui, 2021a), UK (Groarke et al., 2020), and USA (Patrick et al., 2020), However a few works compared situations before and after COVID-19 outbreak, the exception being Cheng et al.'s (2024) study.

The remainder of this paper is organized as follows. Section 2 describes the study data. Section 3 proposes testable hypotheses and explains the empirical method. Section 4 presents the estimation results and their interpretations. Section 5 discusses the results. Section 6 concludes the paper.

2. Data and survey method

In 2015, 2016, and 2017 before the COVID-19 pandemic and in 2021 and 2023 after the pandemic, we conducted internet surveys by sending questionnaires to the same participants. Subsequently, a panel dataset was constructed. However, some participants dropped out of the surveys. New participants were recruited to maintain the sample size for subsequent surveys. Therefore, the structure of the sample is unbalanced.

When we planned the initial survey, we selected the Nikkei Research Company (NRC) to be in charge of conducting the survey because the company has a lot of experience in conducting academic surveys, and the cost is lower than that of other companies. Surveys were were conducted until a sufficient sample was collected. In the survey in 2016, we aimed to gather a sufficient sample size, and the survey was conducted until 10,000 observations were collected. Later, NRC continued to conduct surveys for pursuing identical subjects to construct the Panel data. For example, in the second survey in 2017, we collected 9,130 observations, with about 75% of respondents of the 2016 survey. We then matched respondents from 2017 to those in 2016. Consequently, 7,107 respondents participated in both surveys, while 2,023 were new participants. Similarly, observations were collected in 2018, 2021, and 2023. Various control variables were included in the estimations. Participants who did not respond to questions regarding the control variables were excluded from the sample used in the estimation. Furthermore, we excluded respondents who remained unmarried during this period. As explained in Section 4, to assess the influence of grandchildren, the sample was restricted to those with at least one child. Accordingly, the sample size used in the regression estimations was reduced to approximately 24,000. The sample was divided into two subsamples in the estimations.

The questionnaires included questions about basic individual characteristics such as gender, birth year, and marital status. Economic aspects included annual household income, job status, and educational background. As key variables in the study, questions were asked about the happiness level that was used as a proxy for SWB and about family structure, such as the number of sisters and brothers, number of daughters and sons, and number of granddaughters and grandsons.

The burden of childrearing is thought to be higher with babies and infants than with adult children. The strength of the data is that the variables regarding the number of children changed with different time points. In particular, the data covered eight years, from 2016 to 2023; therefore, participants were likely to have had a child or grandchild during this period. Figure 2 compares the average number of respondents' grandchildren before and after the COVID-19 outbreak. It shows that the number of grandchildren increased significantly regardless of the gender of respondents. Similarly, regarding the average number of children, Figure 3 shows that the number of daughters and sons increased regardless of the gender of respondents, although the difference between the periods was not statistically significant. The effect of newborns can be examined using panel data. Further, the data were collected before and after the COVID-19 outbreak. Hence, we can examine how the effect of children on SWB has changed by considering the impact of the COVID-19 pandemic. Unfortunately, we did not ask the birth year or age of the child or grandchildren in the questionnaire. However, we were able to scrutinize the effects of infants born during the study period on SWB.

COVID-19 restricted human behavior and reduced choices in daily life. Figure 4 indicates that SWB after the pandemic began was significantly lower than before the pandemic. Regardless of the cohort, women generally showed a higher SWB, which is consistent with a previous study that compared SWB between genders (Mitsuyama & Shimizutani, 2019). Individuals in the younger cohort, who were born after 1970, showed a lower SWB than those in the older cohort. Stress from drastic changes in circumstances such as in the workplace, work style, and childrearing were more likely to be higher for younger people because younger people were more likely to be active workers or raise children than the older people. Interestingly, in the younger cohort, women's SWB after the pandemic began was almost equal to men's SWB before the pandemic. This implies that gender differences in SWB are equivalent to the magnitude of the impact of COVID-19 on women's SWB. Assuming men are more likely to be full-time workers, the impact of stress from their workplace on their SWB is similar to that from stress from COVID-19 on women's SWB. However, the degree of reduction in SWB of women was greater than that of men. One reason for this is that the degree of circumstance change may be larger for women than for men. For example, unexpected school closures increased the burden of childrearing more for mothers of school-age children than for fathers (Yamamura & Tsutusi, 2021a). School closure reduced mothers mental health, whereas the mental health of the fathers was less likely to decline (Yamamura & Tsutusi, 2021b).

Table 1 provides definitions of the key variables, their mean values, and standard errors. The SWB and family structure were quantified using these variables. In particular, gender differences and matches within families were captured. Family structure at the time of the surveys and in childhood (number of older and younger sisters and brothers) was considered.

## 3. Hypothesis and method

### 3.1. Hypothesis

Hansen (2012) argued that having a child exerts both positive and negative influences on parents' SWB. Hence, the overall effect of having a child depends on whether the positive effect outweighs the negative effect. Empirical studies found that presence of children reduced their parents' SWB (Blanchflower & Clark, 2021; Margolis & Myrskylä, 2011, Stanca, 2012). The negative relationship might be due to the cost of childrearing being higher than the delight of having a child. If grandparents participate in rearing grandchildren, grandchildren reduce grandparents' SWB (Brunello & Rocco, 2019). Thus, newborn babies trigger bargaining between their parents and grandparents for the division of labor.

The classical work of Becker (1981) proposes, "even small differences between men and women-presumably related at least partially to the advantages of women in the birth and rearing of children-would cause a division of labor by gender, with wives more specialized to household activities and husbands more specialized to other work." Apart from this, the gender identity also causes wives to be burdened with household chores, even if both the wife and husband have full-time work (Akerlof & Kranton, 2000). Inevitably, the difference in SWB between wives and husbands increases after marriage (Bethmann & Rudolf, 2018).

Husbands and wives seem to learn how to live from their parents during childhood. Naturally, the gender identity may be inherited from parents (Yamamura & Tsutsui, 2021c). Researchers have observed differences in the effects of grandchildren on grandmothers and grandfathers and those of children on mothers and fathers (Brunello & Yamamura, 2023; Yamamura & Brunello, 2023).

During the COVID-19 pandemic in Japan, people's activities were restricted to avoid the spread of the virus. The reduction in interpersonal exchanges reduced the degree of grandparents' participation in rearing grandchildren. Thus, the cost of rearing grandchildren reduced for grandparents, whereas the joy of having grandchildren persisted. By contrast, the reduction in grandparents' support increased parents' childrearing costs[2]. These effects are thought to be stronger for grandmothers and mothers than for grandfathers and fathers because childrearing costs are higher for women than for men.

From discussion as above, we propose *Hypothesis 1:*

*Hypothesis 1: Reduction of burden of grandchild rearing has mitigated the negative effect of grandchildren on grandparents' SWB, whereas increase of burden of child rearing has increased the negative effect of children on parents' SWB. The effect is stronger in women.*

Childhood circumstances form a person's perception of and skills in interpersonal relationships through interactions with family members. For instance, the attitude toward household chores is learned from parents' intrahousehold division of labor (Yamamura & Tsutsui, 2021c). The presence of siblings is also important. Playing with younger siblings shapes the skills required to deal with children. Furthermore, gender differences seem to exist even in the case of children. For instance, girls tend to prefer playing house to outdoor recreation, although many do not fit this stereotype. Careful attention should be paid to this simplification. Therefore, skills should be specified according to the gender. That is, the experience of playing with the younger sister builds the skill of rearing a daughter, whereas playing with the younger brother shapes the skill of rearing a son.

---

[2] The cost to have a child includes not only monetary, but also degree of effort and time spent on childrearing.

Furthermore, gender matching between parents and their siblings is critical. The experience of playing at home accumulates among sisters because they share preferences. This also holds true for the relationship among brothers. Thus, we propose *Hypothesis 2*

> *Hypothesis 2: Even after the spread of COVID-19, the SWB of mothers having younger sisters is less likely to reduce even if they have daughters, and that of fathers having younger brothers is less likely to reduce even if they have sons.*

3.2. Method

To examine the influence of grandchildren, the sample should be restricted to those who could become grandparents. Therefore, in the estimations, we split the sample according to birthyear into before and after 1970. To estimate grandchildren's effect, we used a subsample of those who had at least one child and were born before 1970. To estimate children's effect, we used a subsample of those born after 1970.

The baseline model assesses the effects of number of grandchildren on grandparents' SWB. The estimated function takes the following form.

$$SWB_{it} = \alpha_0 + \alpha_1\ GRAND\ CHILD_{it}*AFTER\ COVID_t + \alpha_2\ GRAND\ CHILD_{it} + \alpha_3\ AFTER\ COVID_t + X'_i B + u_i + e_{it} \qquad (1)$$

*SWB $_i$* is the dependent variable. α denotes the coefficients of the variables. *i* and *t* represent individuals and time points, respectively. X is the vector of the control variables, and B is the vector of their coefficients. X represents marital status, age, and household income. The FE model controls an individual's time-invariant characteristics, represented by u $_i$. e $_{it}$ is an error term. The key independent variable is *GRAND CHILD *AFTER COVID*, which is the interaction term between *GRAND CHILD* and *AFTER COVID.*

In the alternative specification after splitting the grandchildren into granddaughters and grandsons, *GRAND CHILD*AFTER COVID* was replaced by *GRAND DAUGHTER*AFTER COVID* and *GRAND SON*AFTER COVID*. To explore the gender match between grandmothers and grandchildren, the sample was further divided into female and male samples.

Similarly, to investigate the influence of the child, the key variable used was the interaction term between *CHILD*AFTER COVID.* In the alternative specification, *CHILD*AFTER COVID* was replaced by *DAUGHTER*AFTER COVID* and *SON*AFTER COVID*.

From *Hypothesis 1*, the signs of the coefficients *GRAND CHILD* and *CHILD* are predicted to be negative. This tendency was expected to be more pronounced in a female sample than in a male sample.

We further investigated how the presence of siblings changed the effects of daughters and sons. As illustrated in Figure 5, relationships with siblings present many possible learning channels to deal with children if we consider gender differences. *Hypothesis 2* states the channels demonstrated by solid arrows that are wider than the dashed ones. Here, we focus on channels that include younger siblings and children of the same gender as the parents. To test *Hypothesis 2*, we used the following specification to control for the effects through all channels shown in Figure 5:

$$SWB_{it} = \alpha_0 + \alpha_1 \text{ ELDER SISTER}_i * \text{DAUGHTER}_{it} * \text{AFTER COVID}_t$$
$$+ \alpha_2 \text{ YOUNGER SISTER}_i * \text{DAUGHTER}_{it} * \text{AFTER COVID}_t$$
$$+ \alpha_3 \text{ ELDER BROTHER}_i * \text{DAUGHTER}_{it} * \text{AFTER COVID}_t$$
$$+ \alpha_4 \text{ YOUNGER BROTHER}_i * \text{DAUGHTER}_{it} * \text{AFTER COVID}_t$$
$$+ \alpha_5 \text{ ELDER SISTER}_i * \text{SON}_{it} * \text{AFTER COVID}_t$$
$$+ \alpha_6 \text{ YOUNGER SISTER}_i * \text{SON}_{it} * \text{AFTER COVID}_t$$
$$+ \alpha_7 \text{ ELDER BROTHER}_i * \text{SON}_{it} * \text{AFTER COVID}_t$$
$$+ \alpha_8 \text{ YOUNGER BROTHER}_i * \text{SON}_{it} * \text{AFTER COVID}_t$$
$$+ X'_i B + V'_i C + u_i + e_{it} \quad (2)$$

Here, the interaction terms among the three variables are the key variables, and we report only their results. Difference from baseline Model (1) is including "V" that represents the vector of variables including other interaction terms consisting of components in key variables. To take an example of *ELDER SISTER*$_i$*DAUGHTER*$_{it}$*AFTER COVID*, we included *ELDER SISTER*DAUGHTER*, *ELDER SISTER*AFTER COVID*, *DAUGHTER*AFTER COVID*, *DAUGHTER* and *AFTER COVID* in "V."

From *Hypothesis 2*, *YOUNGER SISTER*DAUGHTER*AFTER COVID* is expected to have a positive sign in the female sample, whereas *YOUNGER BROTHER*SON*AFTER COVID* is expected to have a positive sign in the male sample.

4. Results and interpretation

Tables 2 and 3 present the scores of the correlation between the number of grandchildren and their grandparents' SWB. We used a subsample of those who had at least one child and were born before 1970 because they might possibly have a grandchild. The results in Table 2 are based on a female sample that reports grandchildren's effect on grandmothers' SWB. The results in Table 3 are from the male sample that reports grandchildren's impact on grandfathers' SWB. Similarly, Tables 4 and 5 report the association between the number of children and parents' SWB. Tables 6 and 7 show the results of Specification (2) proposed in Section 3 to examine how siblings influence the correlation.

4.1. Influence of grandchildren on grandparents

Table 2 shows a significant positive sign of GRAND DAUGHTER*AFTER COVID in both Columns (3) and (4). This implies that granddaughters improved the SWB of their grandmothers after the COVID-19 outbreak than before. The results of GRAND SON*AFTER COVID are not robust, although they show significant negative signs in Column (4). Therefore, the results varied according to the specifications. In Table 3, we do not see robust results for any of the independent variables.

The combined results of Tables 2 and 3 lead us to argue that the burden of rearing grandchildren was reduced after the pandemic began. Therefore, the grandchildren's effect became positive for grandmothers if the gender matched. However, grandfathers' contribution to rearing grandchildren was very small even before the pandemic; therefore, it did not change even after the pandemic. Therefore, grandchildren's influence was the same on grandfathers before and after the pandemic.

4.2. Influence of children on SWB of their parents

As shown in Table 4, none of interaction terms showed statistical significance. Therefore, the influence of children on mothers' SWB did not change even after the pandemic, which is inconsistent with the short-term analysis (Yamamura & Tsutsui, 2021b).

By contrast, Table 5 indicates a significant negative sign of CHILD*AFTER COVID in Columns (1) and (2). Even after dividing the children by gender, significant negative signs were observed in the DAUGHTER*AFTER COVID and SON*AFTER COVID groups. Therefore, both daughters and sons reduced their fathers' SWB after the outbreak. Interestingly, SON showed a positive sign and statistical significance, whereas no statistical significance was observed in DAUGHTER. Therefore, before the pandemic, sons increased their fathers' SWB, whereas daughters did not. The combined results of Tables 4 and 5 imply that same-gender children are preferred by men, but not by women. This might be partly because the husband was unlikely to be burdened with childrearing, and he only enjoyed with children. In compared with opposite gender, same gender has similar preferences. Thus, there is no need for the cost of learning about the same gender's preference. Naturally, the net benefit from the son to the father is greater than that from the daughter. In other words, the husband's utility increased with "preferred goods (sons)".

The results for grandparents largely supported *Hypothesis 1*. However, the finding that children reduced their fathers' SWB but not their mothers' SWB was inconsistent with *Hypothesis 1*. This suggests that the pandemic may have changed the father's role in the family. Fathers' time at home became longer than before the pandemic, leading them to spend more time with their children. Therefore, the allocation of the burden of childrearing is considered to have changed between wives and husbands and an increase in the fathers' burden reduced their SWB.

4.3 Results on effects of siblings

Tables 6 and 7 report only the key interaction variables consisting of three components. Other interaction variables consisting of two components and other control variables used in Tables 5 and 6 are included as independent variables.

In Table 6, Columns (1) and (2) indicate significant positive signs for ELDER SISTER*DAUGHTER*AFTER COVID. This implies that the SWB of mothers with older sisters and daughters was more likely to be higher than that of mothers with daughters after the pandemic. Columns (3) and (4) show positive signs for YOUNGER BROTHER*SON*AFTER COVID and the values are statistically significant at the 1% level. After the pandemic, the SWB of fathers with younger brothers and sons was more likely to be higher than that of fathers with sons. These results indicate that the SWB of men and women with children and siblings of the same gender as theirs is more likely to have been higher than that of others after the pandemic.

For mothers, learning from elder sisters develops their skills in dealing with younger sisters from the viewpoint younger sisters. This skill can be useful for daughter rearing. For fathers, there may be different mechanisms by which younger brothers improve their elder brothers' skill of playing with a small boy. This skill can be directly applied to son rearing. Women are more mature than men in childhood, so learning from elder sisters is greater than that from younger sisters. In other words, men are less mature, so they may learn directly from their own experience as an elder brother, not from how their elder brothers deal with them.

Table 7 shows a point difference from Table 6 in that the number of siblings was replaced by a sibling dummy, as defined in Table 1. The results in Table 7 are similar to those in Table 6. However, the linear combination of the interaction terms yields additional results. As shown in Table 5, sons reduced their fathers' SWB after the pandemic. However, Table 7 indicates that the presence of younger brothers mitigates this negative effect. Here, we considered the combined results using a linear combination. As clearly shown in Columns (3) and (4), the results of the linear combination are negative but not statistically significant. Thus, fathers with younger brothers exhibited the same SWB before and after the pandemic because younger brothers' presence may have neutralized sons' negative effect on SWB. In our interpretation, learnings through experiences in the family in childhood about to deal with small boys are critical for fathers' childrearing in an emergency.

*Hypothesis 2* was supported by the results of fathers' SWB but not by mothers' SWB. Due to differences in childhood maturity, the effects of siblings differed. Men who played the father's role as elder brothers in childhood seemed to deal well with their sons. Women who learned from elder sisters and mimicked them in childhood could deal well with their daughters.

5. Discussion

The contribution of this study is that it is the first to provide evidence on the effect of children on the SWB of parents and grandparents by comparing their SWB before and after the COVID-19 outbreak. The evidence has policy implications for childrearing post COVID-19.

Granddaughters increased their grandmothers' SWB. In our interpretation, COVID-19

reduced grandmothers' burden of childcare to avoid infection, although the burden on grandfathers did not change because they did not play a substantial role in it even before COVID-19. Reduction in the cost of childrearing increased the net benefit from grandchildren. By contrast, both daughters and sons reduced their fathers' SWB but not their mothers'. This reflects the fact that an increase in the cost of childrearing reduced fathers' net benefit from children. However, the mothers were better able to adapt to the change, so their SWB did not differ from that before COVID-19.

There are two possible interpretations of the reduction in fathers' SWB. First, as we already argued in the previous section, with the spread of COVID-19, men were more likely to have spent time on childrearing, and an increase in the cost of childrearing reduced their SWB. The second interpretation is possibly a spurious correlation, and a hidden mechanism might exist as follows. Assume that husbands did not substantially change their lifestyle to spend a long time in the workplace (Yamamura & Tsutsui, 2021b), and only wives spent more time with children than before the pandemic. Consequently, the gap in time spent interacting with children widened between wives and husbands, and the relationship between mothers and children strengthened. Inevitably, it became difficult for fathers to find their place in the house if they returned late. Fathers were thus isolated from their family, which reduced their SWB.

Our finding that the presence of younger brothers reduces sons' negative effect on fathers' SWB leads us to argue that fathers were more likely to participate in childrearing. Therefore, the second conjecture would not hold. We could argue that husbands came to perceive the stress from childrearing during the pandemic and attempted to balance work and childrearing. This way, they encountered difficulties that their wives had experienced for many years. Post COVID-19, the mutual understanding between the wife and husband would be enhanced because they share the childrearing problem.

Participating in household chores and childrearing might form a non-cognitive skill required in the post-pandemic society. It seems useful for children to play together at home during their childhood regardless of their gender because they can learn to share household chores as well as childrearing. While playing, small children imagine future family life situations. This is an image training for appropriately coordinating family life. For instance, the boy could imagine the cost of childrearing and household chores through an interaction with his spouse if he participated in play with the girl. The number of children without siblings increased because of the drop in the birth rate. This hampers the development of skills for dealing with children, further reducing birth rates. To break this vicious circle, children should experience having siblings outside their families. The policy implication of these findings is that playing house should be promoted in nursery schools and kindergartens.

The limitations of this study are as follows. The way of working and job status might have changed during the study period, which may have different effects on SWB. Individuals' SWB may depend on whether they live with their children (grandchildren). However, owing to the limitations of the dataset, we could not scrutinize how the effect of family members differs according to whether they live together. Furthermore, the situation cannot simply be divided into having a child and not having one. During the study period, respondents or respondents' spouses were likely to be pregnant. Behavior during pregnancy differs from both before and after having a child (Yamamura & Tsutsui, 2019), seemingly influencing SWB (Hagstrom & Wu, 2016). These issues should be addressed in future studies.

# 6. Conclusion

After the COVID-19 pandemic began, the difference in the burden of childrearing has widened between mothers and fathers, leading mothers' SWB to decline (Yamamura & Tsutsui, 2021a, 2021b). Approximately five years have passed since the COVID-19 outbreak in 2019. As we enter the post-COVID world, we naturally ask the question: Does the influence of COVID-19 persist in daily life? The impact of various life events such as marriage and divorce on SWB is temporary (Clark et al., 2008). However, little is known about how COVID-19 has altered family relations to influence SWB in the long term.

The COVID-19 pandemic is considered to have detached grandparents from rearing grandchildren. The research questions of this study were as follows: How does the change in family relations impact the SWB of grandparents and parents?

Using individual-level panel data over 2016–2023, we found that (1) granddaughters were positively correlated with their grandmothers' SWB after COVID-19, (2) both daughters and sons were negatively correlated with their fathers' SWB after COVID-19, whereas neither daughters nor sons were associated with their mothers' SWB, and (3) the negative correlation between sons and their fathers' SWB was not observed if their fathers had younger brothers.

The experience of playing with siblings during childhood forms a non-cognitive skill for dealing with children. Both girls and boys should be encouraged to play together at home. Games like roleplay to have siblings are important because the number of children who do not have siblings will increase because of a persistent drop in birthrate. To break the vicious circle of population decline, promoting playing house might be useful in building non-cognitive skills in nursery schools and kindergartens. However, little is known about the effectiveness of playing at home in childhood in shaping the skills of rearing children and doing household chores in adulthood. This should be investigated in future studies.

## Statements and declarations

There is no conflict of interest.

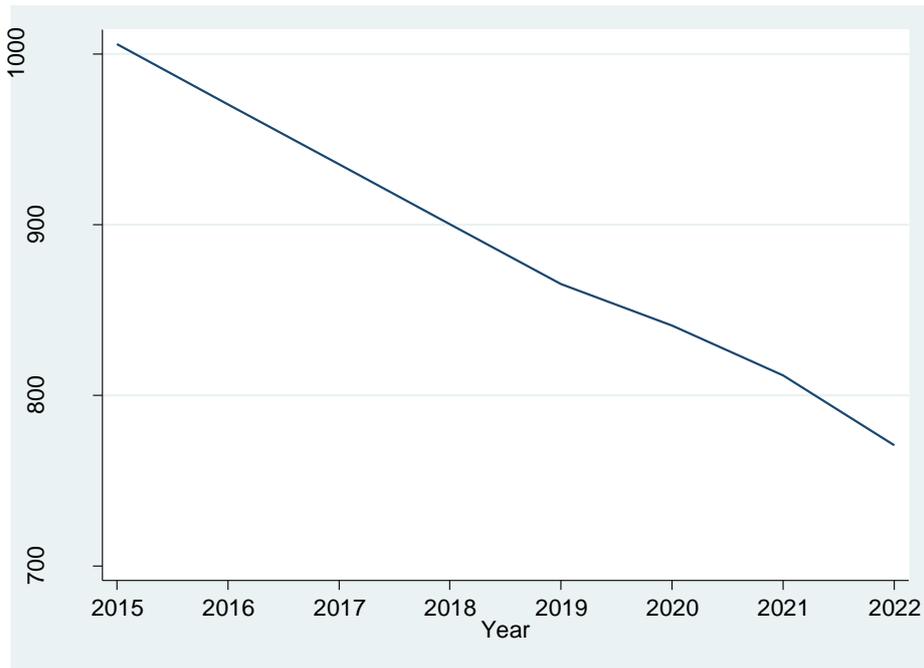

**Fig. 1** Number of births in Japan

Source: Ministry of Health, Labour and Welfare.

https://www.mhlw.go.jp/toukei/saikin/hw/jinkou/geppo/nengai22/index.html　Accessed on Nov 11, 2023.

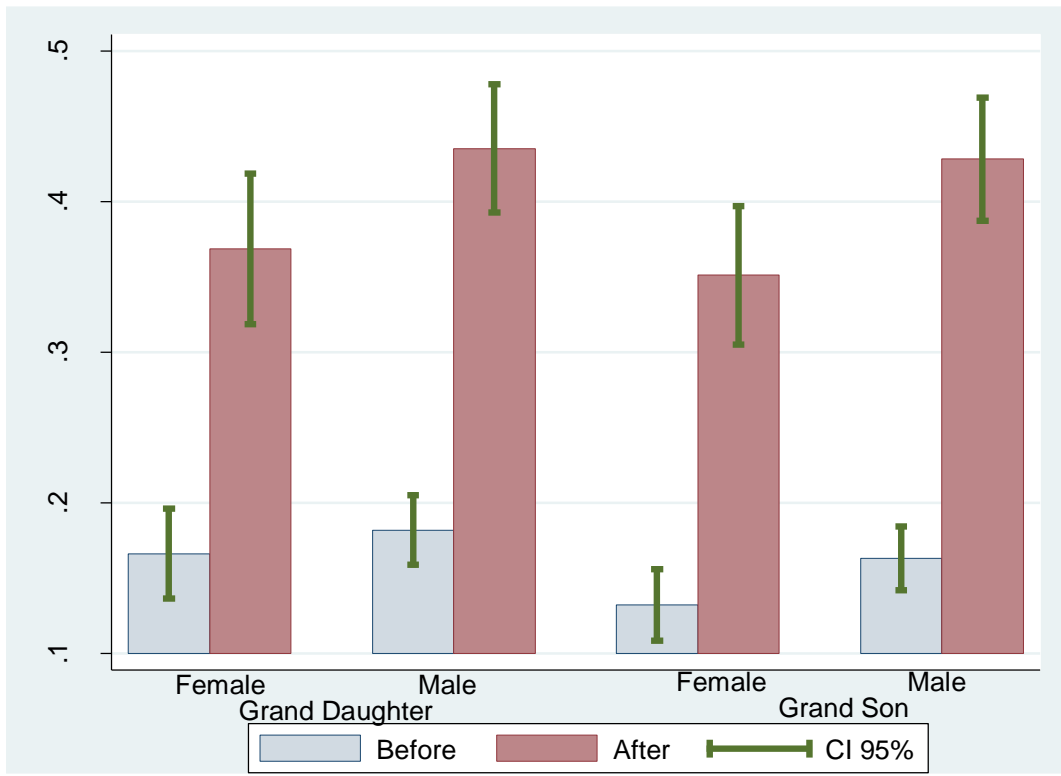

**Fig. 2** Number of grandchildren before and after COVID-19 outbreak (Respondents' birth year <=1970)

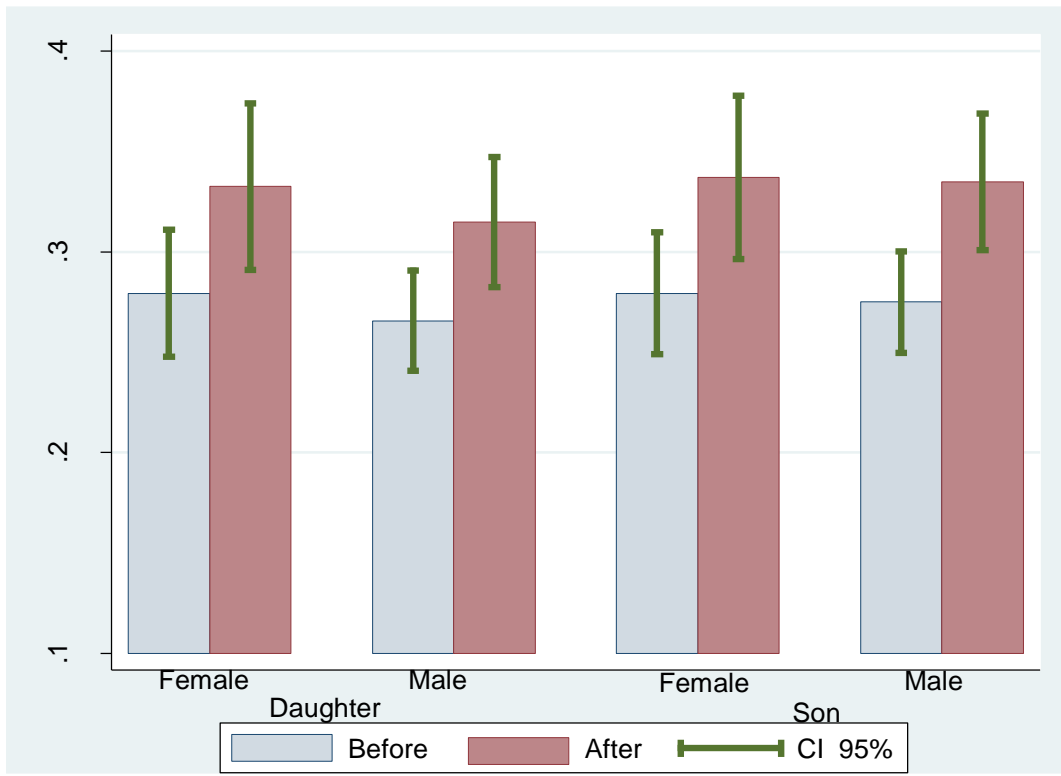

**Fig. 3** Number of children before and after COVID-19 outbreak (Respondents' birth year >1970)

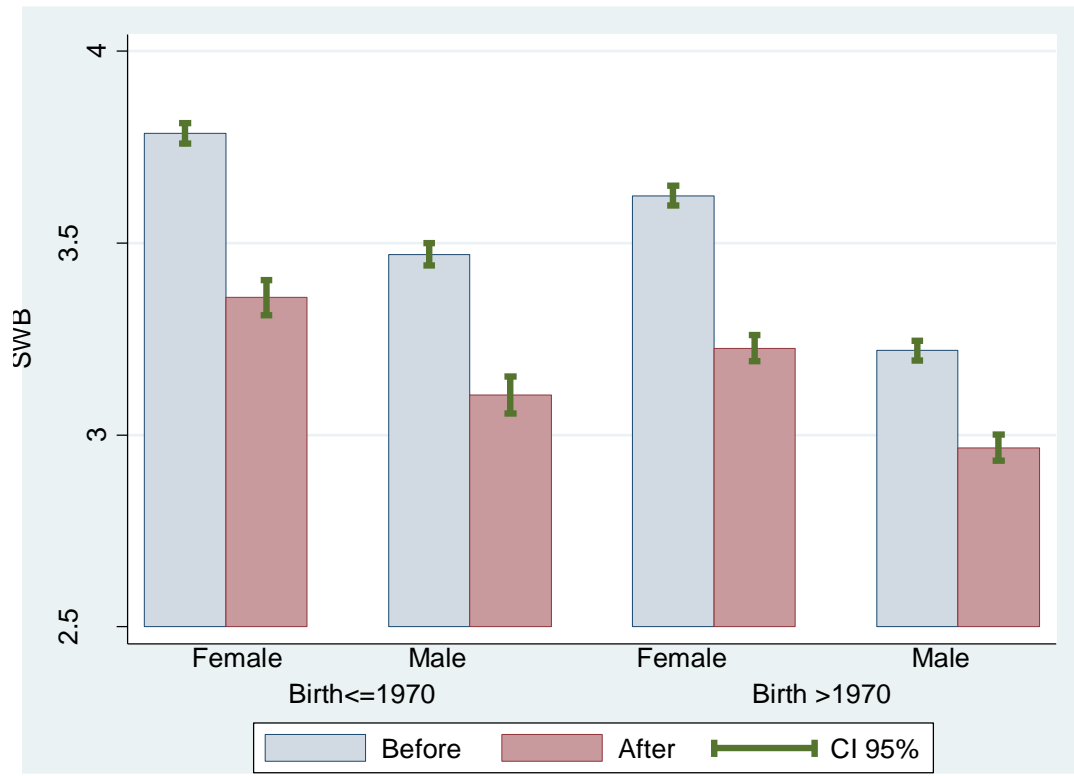

**Fig. 4** Change in SWB before and after COVID-19 outbreak

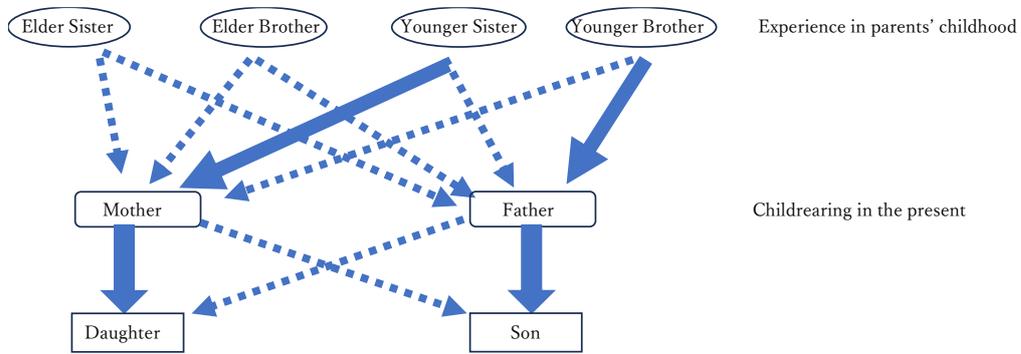

**Fig. 5** Channels of learning skills through genders matches within a family

Table 1. Definitions of variables and their basic statistics: Sample restricted to those who have at least one child.

| Variables | Definition | Mean | S.D. |
|---|---|---|---|
| SWB | Are you happy? 1 (Unhappy) – 5 (Happy) | 3.39 | 1.18 |
| GRAND CHILD | Number of grandchildren | 0.34 | 1.08 |
| GRAND DAUGHTER | Number of granddaughters | 0.17 | 0.57 |
| GRAND SON | Number of grandsons | 0.17 | 0.58 |
| CHILD | Number of children | 1.07 | 1.17 |
| DAUGHTER | Number of daughters | 0.52 | 0.77 |
| SON | Number of sons | 0.55 | 0.79 |
| AFTER COVID | Equals 1 for the survey conducted after 2021 or 2023, 0 otherwise | 0.30 | 0.46 |
| ELDER SISTER | Number of elder sisters | 1.37 | 1.74 |
| YOUNG SISTER | Number of younger sisters | 1.99 | 2.60 |
| ELDER BROTHER | Number of elder brothers | 1.36 | 1.75 |
| YOUNG BROTHER | Number of younger brothers | 1.42 | 1.71 |
| ELDER SISTER D | Equals 1 if respondent has elder sister, 0 otherwise | 0.47 | 0.49 |
| YOUNG SISTER D | Equals 1 if respondent has younger sister, 0 otherwise | 0.51 | 0.49 |
| ELDER BROTHER D | Equals 1 if respondent has elder brother, 0 otherwise | 0.45 | 0.49 |
| YOUNG BROTHER D | Equals 1 if respondent has younger brother, 0 otherwise | 0.52 | 0.49 |

Table 2. Effect of grandchildren on grandmothers' SWB. Respondent has at least one child and her birth year<=1970. (Fixed effects model)[a]

|  | (1) Full sample | (2) Participate all | (3) Full sample | (4) Participate all |
|---|---|---|---|---|
| GRAND CHILD * AFTER COVID | −0.001 (−0.03) | −0.039 (−1.24) |  |  |
| GRAND CHILD | −0.008 (−0.36) | 0.017 (0.58) |  |  |
| GRAND DAUGHTER * AFTER COVID |  |  | 0.058* (1.78) | 0.064** (1.94) |
| GRAND DAUGHTER |  |  | −0.039 (−1.28) | −0.034 (−0.81) |
| GRAND SON * AFTER COVID |  |  | −0.066 (−1.38) | −0.174*** (−2.62) |
| GRAND SON |  |  | 0.033 (0.76) | 0.099 (1.59) |
| AFTER COVID | 0.346*** (4.80) | 0.224** (2.39) | 0.349*** (4.85) | 0.236** (2.53) |
| Within R-squared | 0.06 | 0.07 | 0.06 | 0.07 |
| Individuals | 2,037 | 349 | 2,037 | 349 |
| Observations | 4,837 | 1,692 | 4,837 | 1,692 |

[a] ***, **, and * denote statistical significance at the 1%, 5%, and 10% levels, respectively. The t-values are calculated based on robust standard errors. *Participate all* means that samples of individuals who participated in all surveys were included, although they were not included in some early surveys before they had a child. Various control variables such as respondents' age, marital status dummies, and household income dummies were included. However, these estimates have not been reported.

Table 3. Effect of grandchildren on grandfathers' SWB. Respondent has at least one child and his birth year<=1970. (Fixed effects model)[a]

|  | (1) Full sample | (2) Participate all | (3) Full sample | (4) Participate all |
|---|---|---|---|---|
| GRAND CHILD * AFTER COVID | 0.007 (0.43) | 0.018 (0.90) |  |  |
| GRAND CHILD | −0.030** (−2.27) | −0.056*** (−2.61) |  |  |
| GRAND DAUGHTER * AFTER COVID |  |  | 0.045 (1.58) | 0.042 (1.24) |
| GRAND DAUGHTER |  |  | −0.051* (−1.77) | −0.060 (−1.64) |
| GRAND SON * AFTER COVID |  |  | −0.032 (−0.96) | −0.008 (−0.18) |
| GRAND SON |  |  | −0.027 (−0.90) | −0.050 (−1.20) |
| AFTER COVID | 0.165*** (2.93) | 0.041 (0.61) | 0.164*** (2.92) | 0.041 (0.61) |
| Within R-squared | 0.06 | 0.07 | 0.06 | 0.07 |
| Individuals | 2,029 | 510 | 2,029 | 510 |
| Observations | 5,365 | 2,486 | 5,365 | 2,486 |

[a] ***, **, and * denote statistical significance at the 1%, 5%, and 10% levels, respectively. The t-values are calculated based on robust standard errors. *Participate all* means that samples of individuals who participated in all surveys were included, although they were not included in some early surveys before they had a child. Various control variables such as respondents' age, marital status dummies, and household income dummies were included. However, these estimates have not been reported.

Table 4. Effect of children on mothers' SWB. Respondent's birth year>1970. (Fixed effects model)[a]

|  | (1) Full sample | (2) Participate all | (3) Full sample | (4) Participate all |
|---|---|---|---|---|
| CHILD * AFTER COVID | −0.053 (−1.61) | −0.026 (−0.76) |  |  |
| CHILD | 0.032 (0.47) | −0.022 (−0.28) |  |  |
| DAUGHTER * AFTER COVID |  |  | −0.041 (−0.98) | −0.012 (−0.27) |
| DAUGHTER |  |  | 0.007 (0.80) | −0.006 (−0.06) |
| SON * AFTER COVID |  |  | −0.061 (−1.31) | −0.039 (−0.80) |
| SON |  |  | −0.019 (−0.18) | −0.040 (−0.31) |
| AFTER COVID | 0.197*** (2.91) | 0.118 (1.43) | 0.197*** (2.91) | 0.119 (1.43) |
| Within R-squared | 0.06 | 0.09 | 0.06 | 0.09 |
| Individuals | 2,913 | 440 | 2,913 | 440 |
| Observations | 6,144 | 2,127 | 6,144 | 2,127 |

[a] ***, **, and * denote statistical significance at the 1%, 5%, and 10% levels, respectively. The t-values are calculated based on robust standard errors. *Participate all* means that samples of individuals who participated in all surveys were included. Various control variables such as respondents' age, marital status dummies, and household income dummies were included. However, these estimates have not been reported.

Table 5. Effect of children on fathers' SWB. Respondent's birth year>1970. (Fixed effects model)[a]

|  | (1) Full sample | (2) Participate all | (3) Full sample | (4) Participate all |
|---|---|---|---|---|
| CHILD * AFTER COVID | −0.088*** (−3.59) | −0.091*** (−3.54) |  |  |
| CHILD | 0.033 (0.54) | 0.023 (0.34) |  |  |
| DAUGHTER * AFTER COVID |  |  | −0.093** (−2.20) | −0.077* (−1.89) |
| DAUGHTER |  |  | −0104 (−1.19) | −0.122 (−1.08) |
| SON * AFTER COVID |  |  | −0.085** (−2.57) | −0.102*** (−2.96) |
| SON |  |  | 0.163** (2.08) | 0.154** (1.99) |
| AFTER COVID | 0.124** (2.34) | 0.085 (1.46) | 0.126** (2.37) | 0.090 (1.53) |
| Within R-squared | 0.04 | 0.06 | 0.04 | 0.06 |
| Individuals | 3,224 | 719 | 3,224 | 719 |
| Observations | 7,907 | 3,538 | 7,907 | 3,538 |

[a] ***, **, and * denote statistical significance at the 1%, 5%, and 10% levels, respectively. The t-values are calculated based on robust standard errors. *Participate all* means that samples of individuals who participated in all surveys were included. Various control variables such as respondents' age, marital status dummies, and household income dummies were included. However, these estimates have not been reported.

Table 6. Impact of interaction with sisters and brothers on SWB. Respondent's birth year>1970. (Fixed effects model)[a]

|  | (1) Mother Full sample | (2) Mother Participate all | (3) Father Full sample | (4) Father Participate all |
|---|---|---|---|---|
| ELDER SISTER*DAUGHTER * AFTER COVID | 0.294** (2.32) | 0.274* (1.93) | −0.045 (−0.43) | −0.011 (−0.12) |
| YOUNGER SISTER*DAUGHTER * AFTER COVID | 0.061 (0.90) | 0.094 (1.24) | −0.036 (−0.45) | −0.057 (−0.72) |
| ELDER BROTHER*DAUGHTER * AFTER COVID | 0.060 (0.63) | 0.040 (0.38) | 0.051 (0.47) | 0.160 (1.65) |
| YOUNGER BROTHER*DAUGHTER * AFTER COVID | 0.070* (1.76) | 0.034 (0.93) | −0.006 (−0.08) | 0.018 (0.22) |
| ELDER SISTER*SON * AFTER COVID | 0.042 (0.57) | 0.034 (0.47) | 0.140* (1.87) | 0.132 (1.45) |
| YOUNGER SISTER*SON * AFTER COVID | 0.004 (0.06) | −0.037 (−0.46) | 0.156* (1.95) | 0.141 (1.63) |
| ELDER BROTHER*SON * AFTER COVID | 0.128 (1.28) | 0.163 (1.37) | 0.151* (1.96) | 0.117 (1.54) |
| YOUNGER BROTHER*SON * AFTER COVID | 0.123 (0.95) | 0.189 (1.36) | 0.173*** (2.64) | 0.199*** (2.90) |
| Within R-squared | 0.06 | 0.09 | 0.04 | 0.06 |
| Individuals | 2,913 | 440 | 3,224 | 719 |
| Observations | 6,144 | 2,127 | 7,907 | 3,538 |

[a] ***, **, and * denote statistical significance at the 1%, 5%, and 10% levels, respectively. The t-values are calculated based on robust standard errors. *Participate all* means that samples of individuals who participated in all surveys were included. Various control variables such as respondents' age, marital status dummies, and household income dummies were included. However, these estimates have not been reported.

Table 7. Interaction terms of dummies for sister and brother and impact on SWB. Respondent's birth year>1970. (Fixed effects model)[a]

|  | (1) Mother Full sample | (2) Mother Participate all | (3) Father Full sample | (4) Father Participate all |
|---|---|---|---|---|
| ELDER SISTER D*DAUGHTER * AFTER COVID | 0.316** (2.33) | 0.287* (1.90) | −0.071 (−0.60) | −0.028 (−0.25) |
| YOUNGER SISTER D*DAUGHTER * AFTER COVID | 0.082 (0.80) | 0.099 (0.88) | −0.100 (−0.94) | −0.092 (−0.89) |
| ELDER BROTHER D*DAUGHTER * AFTER COVID | 0.036 (0.27) | −0.007 (−0.05) | 0.020 (0.15) | 0.171 (1.58) |
| YOUNGER BROTHER D*DAUGHTER * AFTER COVID | 0.098 (1.04) | 0.036 (0.35) | −0.078 (−0.65) | −0.034 (−0.29) |
| ELDER SISTER D*SON * AFTER COVID | −0.001 (−0.01) | −0.008 (−0.09) | 0.133 (1.33) | 0.119 (1.14) |
| YOUNGER SISTER D*SON * AFTER COVID | −0.026 (−0.30) | −0.060 (−0.64) | 0.155* (1.84) | 0.137 (1.54) |
| ELDER BROTHER D*SON * AFTER COVID | 0.119 (1.27) | 0.220** (2.20) | 0.152 (1.39) | 0.140 (1.26) |
| YOUNGER BROTHER D*SON * AFTER COVID | 0.075 (0.61) | 0.192 (1.55) | 0.239*** (2.61) | 0.264*** (2.74) |
| Linear combination |  |  |  |  |
| SON * AFTER COVID+ YOUNGER BROTHER D*SON * AFTER COVID | −0.033 (−0.35) | 0.063 (0.63) | −0.042 (−0.74) | −0.040 (−0.66) |
| Within R-squared | 0.07 | 0.10 | 0.05 | 0.07 |
| Individuals | 2,913 | 440 | 3,224 | 719 |
| Observations | 6,144 | 2,127 | 7,907 | 3,538 |

[a] ***, **, and * denote statistical significance at the 1%, 5%, and 10% levels, respectively. The t-values are calculated based on robust standard errors. *Participate all* means samples of individuals who participated in all surveys were included. Various control variables such as respondents' age, marital status dummies, and household income dummies were included. However, these estimates have not been reported.